\pdfoutput=1
\documentclass{jpsj-suppl}
\usepackage{bm,txfonts} 

\title{Studying short-range correlations and momentum distributions with unitarily transformed operators}

\author{Thomas \textsc{Neff}$^{1}$, Hans \textsc{Feldmeier}$^{1,2}$, Wataru \textsc{Horiuchi}$^{3}$ and Dennis \textsc{Weber}$^{1}$}

\inst{$^{1}$GSI Helmholtzzentrum f{\"u}r Schwerionenforschung GmbH and ExtreMe Matter Institute EMMI, Planckstra{\ss}e 1, 64291 Darmstadt, Germany\\
$^{2}$Frankfurt Institute for Advanced Studies, Max-von-Laue Stra{\ss}e 1, 60438 Frankfurt, Germany\\
$^{3}$Department of Physics, Hokkaido University, Sapporo 060-0810, Japan}

\email{t.neff@gsi.de}

\abst{Short-range correlations in $^4$He are investigated using many-body wave functions obtained in the no-core shell model. The similarity renormalization group (SRG) is used to evolve the Argonne~V8' interaction and the density operators. The effects of short-range correlations are reflected in the two-body densities in coordinate space as a function of the distance between two nucleons, or alternatively in in momentum space as function of the relative momentum between two nucleons. The SRG transformation is performed in two-body approximation. The importance of missing three-body and higher-body contributions is investigated by comparing results obtained for different flow parameters and by comparing to exact results with the bare Argonne~V8' interaction obtained in the correlated Gaussian approach.
}

\kword{short-range correlations, unitary transformations, SRG, NCSM}

\newcommand{\op}[1]{\hat{#1}}
\newcommand{\adj}[1]{{#1}^\dagger}
\renewcommand{\vec}[1]{\bm{\mathit{#1}}}

\newcommand{\comm}[2]{\ensuremath{\left[{#1},{#2}\right]}}

\newcommand{\ket}[1]{\ensuremath{\,|{#1}\rangle}}

\newcommand{\matrixe}[3]{\ensuremath{\langle{#1}|\,{#2}\,|{#3}\rangle}}

\newcommand{\fm}{\ensuremath{\textrm{fm}}}

\newcommand{\nuc}[2]{$^{#1}${#2}}

\begin{document}
\maketitle

\section{Introduction}

Short-range correlations in nuclei reflect properties of the nucleon-nucleon interaction at short distances or high momentum transfers. JLAB experiments \cite{subedi08} measuring ($e$, $e'pn$) and ($e$, $e'pp$) knockout of nucleon pairs at high energies found for example a strong dominance of $pn$- over $pp$-pairs. This has its origin in the strong tensor force that induces corresponding strong tensor correlations in the nuclear many-body wave function. Such short-range correlations are however difficult to describe in many-body approaches. Jastrow-type correlation functions are used in variational or Green's function Monte-Carlo \cite{forest96,wiringa14} and correlated basis function \cite{alvioli13} methods. In these approaches the short-range correlations are incorporated into the wave functions explicitly. Another popular approach is to use unitary transformations to soften the interaction as in the unitary correlation operator method (UCOM) \cite{ucom03,ucom10} or the similarity renormalization group (SRG) \cite{bogner03,bogner10}. The unitary transformations are typically not applied to the wave functions but mapped onto the Hamiltonian. The wave functions obtained with such a transformed Hamiltonian will not have the strong short-range correlations as obtained with the original interaction. An important point here is that also all other observables should be transformed in the same way as the Hamiltonian. This has been studied in detail for the two-body system in \cite{anderson10}. However, many low-energy observables are not really sensitive to the short-range behavior of the wave function and can be calculated in good approximation with the untransformed operators. For other observables that are sensitive to short-range correlations like momentum distributions it is however essential to use the transformed operators.

In this contribution we study short-range correlations in \nuc{4}{He} with the Argonne~V8' interaction \cite{wiringa95} by analyzing two-body densities in coordinate and momentum space. Our calculations use many-body wave functions obtained in the no-core shell model (NCSM) \cite{barrett13} with an SRG transformed Hamiltonian. By comparing the calculated two-body densities obtained with the bare and the SRG transformed density operators with exact results for the bare interaction obtained with correlated Gaussians \cite{src11} we can illustrate how the SRG transformations indeed remove the short-range correlations from the wave functions and how the short-range correlations can be recovered by using transformed density operators.

\section{Method}

The SRG transformed Hamiltonian $\op{H}_\alpha$ and the SRG transformation matrix $\op{U}_\alpha$ are obtained by solving on a momentum space grid the SRG flow equation in two-body space for the Hamiltonian and the transformation matrix:
\begin{equation}
  \frac{d\op{H}_\alpha}{d\alpha} = \comm{\op{\eta}_\alpha}{\op{H}_\alpha}, \hspace{3em}
  \frac{d\op{U}_\alpha}{d\alpha} = - \op{U}_\alpha \op{\eta}_\alpha, \hspace{3em}
  \op{\eta}_\alpha = (2\mu)^2\; \comm{\op{T}}{\op{H}_\alpha} \: .
\end{equation}
In case of the Argonne~V8' interaction relative momenta up to $k_\mathrm{max} = 15\,\fm^{-1}$ are included. The standard meta generator $\op{\eta}_\alpha$ given by the commutator of the kinetic energy and the Hamiltonian is employed. By restricting the evolution to the two-body space the effective Hamiltonian is obtained in two-body approximation where higher order contributions are omitted
\begin{equation}
 \op{H}_\alpha = \adj{\op{U}_\alpha} \op{H} \op{U}_\alpha = \op{T} + \op{V}_\alpha^{[2]} + \ldots + \op{V}_\alpha^{[N]} \approx \op{T} + \op{V}_\alpha^{[2]} \: .
\end{equation}
This transformation has to be performed for other operators as well. Due to the two-body approximation the SRG transformation is unitary only in the two-body space. For larger systems the result with the SRG transformed Hamiltonian will be different from those obtained with the original Hamiltonian. The role of missing higher order contributions can be estimated by analyzing the dependence of the results on the value of the SRG flow parameter $\alpha$. If higher-order contributions are small, there should be only a weak $\alpha$-dependence. 

For the calculation of the two-body densities we first solve the many-body problem with the effective Hamiltonian using the NCSM 
\begin{equation}
 \op{H}_\alpha \ket{\Psi_\alpha} = E_\alpha \ket{\Psi_\alpha} \: .
\end{equation}
With the obtained wave functions $\ket{\Psi_\alpha}$ the bare and SRG transformed two-body densities (in two-body approximation) can then be calculated:
\begin{equation}
 \rho_{\textrm{bare}} = \matrixe{\Psi_\alpha}{\op{\rho}}{\Psi_\alpha}, \hspace{3em}
      \rho_{\textrm{eff}} = \matrixe{\Psi_\alpha}{\adj{\op{U}}_\alpha \, \op{\rho} \, \op{U}_\alpha}{\Psi_\alpha} \: .
\end{equation}

\section{Results}

\begin{figure}
  \centering
  \includegraphics[width=0.83\textwidth]{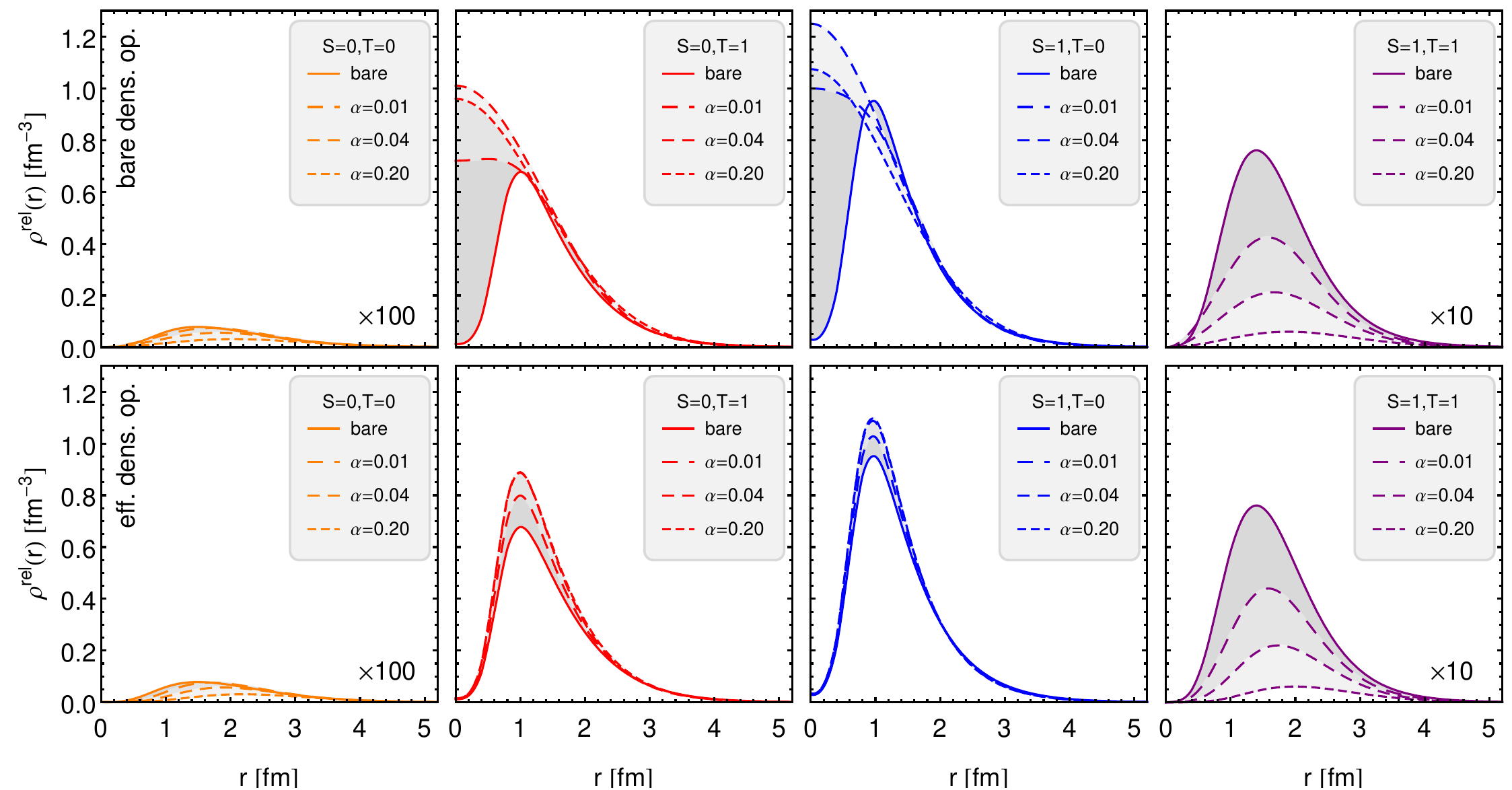}
  \caption{Two-body densities in coordinate space for the different spin-isospin channels in \nuc{4}{He} calculated with NCSM wave functions and an SRG evolved Argonne~V8' interaction with flow parameters of $\alpha = 0$ (bare), $\alpha = 0.01\,\fm^4$, $\alpha = 0.04\,\fm^4$ and $\alpha = 0.20\,\fm^4$. On the top the two-body densities calculated with bare density operators, on the bottom with SRG evolved density operators.}
  \label{fig:He4densR}
\end{figure}
\begin{figure}
  \centering
  \vspace{-2ex}
  \includegraphics[width=0.83\textwidth]{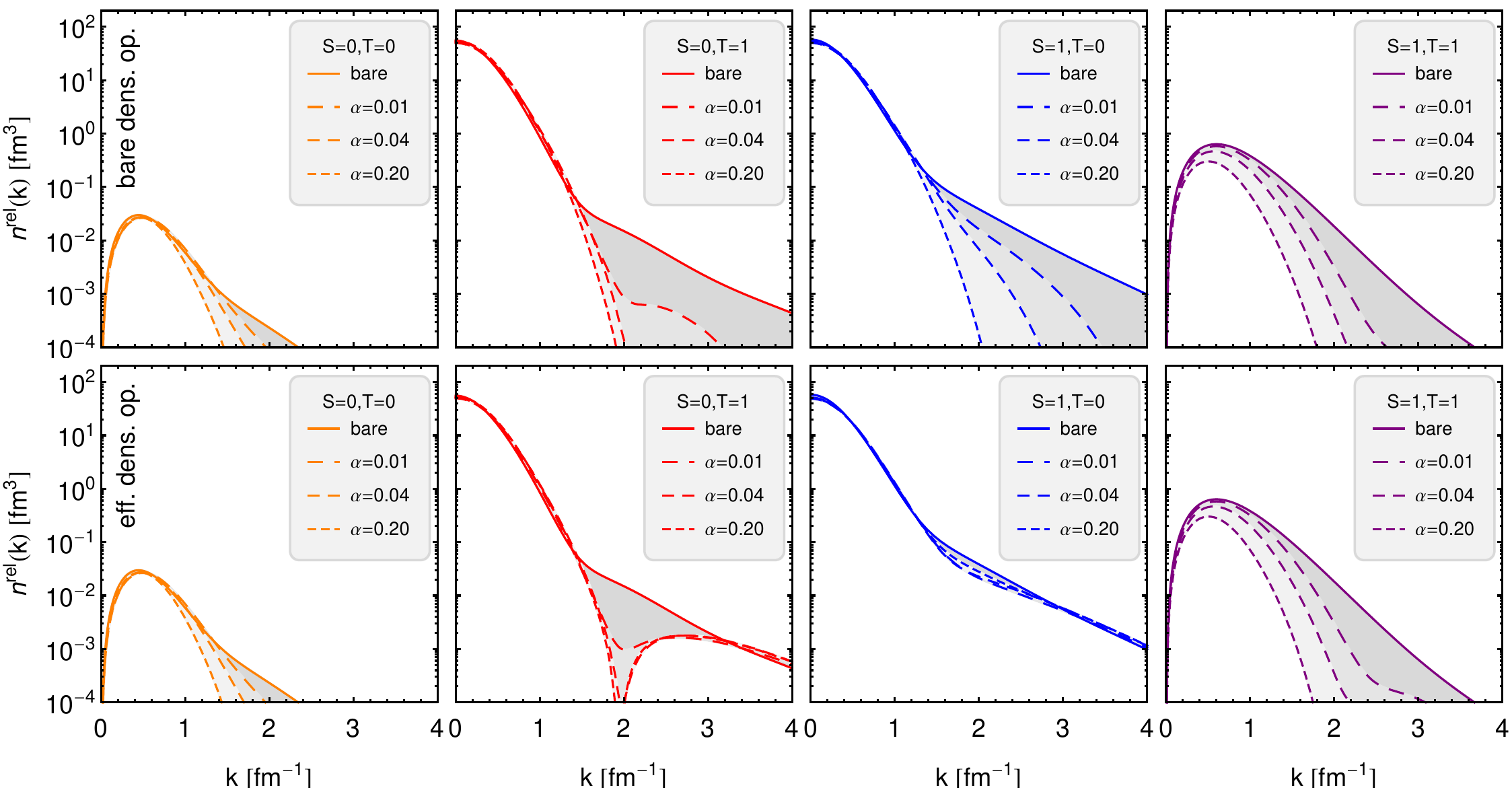}
  \caption{The same as Fig.~\ref{fig:He4densR} but for the two-body densities in momentum space.}
  \label{fig:He4densQ}
\end{figure}

In Fig.~\ref{fig:He4densR} the two-body densities in coordinate space have been calculated for \nuc{4}{He} as a function of the distance of the nucleons and for the different spin-isospin channels:
\begin{equation}
 \op{\rho}^{\mathrm{rel}}_{ST}(\vec{r}) = \sum_{i<j} \delta^{(3)}(\op{\vec{r}}_i-\op{\vec{r}}_j - \vec{r}) \, \op{\Pi}^{ST}_{ij} \: .
\end{equation}
The NCSM calculations with the SRG evolved interactions are well converged within the $N_\mathrm{max}=16$ model space. However with the bare Argonne~V8' interaction the NCSM calculations are not converged and we use here results obtained with the correlated Gaussian method \cite{mitroy13,src11}. As expected the two-body densities calculated with the bare density operators (top panel) show a very strong $\alpha$-dependence. The two-body densities clearly show the correlation hole induced by the repulsive core in the Argonne interaction. For the SRG evolved interactions the correlation hole can no longer be observed -- the short-ranged repulsive core of the interaction has been transformed away. For the two-body densities calculated with the transformed density operators (bottom panel) we would expect the results to be only weakly dependent on the SRG flow parameter $\alpha$ if the two-body approximation is justified. This is indeed the case when we look at the even channels, especially at short distances. The 
calculations with transformed Hamiltonians and transformed density operators show the same correlation hole as obtained with the bare interaction and bare density operators. A certain $\alpha$-dependence can be seen at distances around 1 fm and in the odd-channels. In \nuc{4}{He} only a small number of pairs is found in the odd channels and they reflect three-body correlations \cite{src11} that have their origin in the two-body tensor force. With increasing flow parameter the tensor force in the effective Hamiltonian gets weakened and the population in the odd channels is reduced. The $\alpha$-dependence in the odd channels therefore directly reflects the role of omitted three-body forces in the transformed Hamiltonian.

The two-body densities in momentum space as a function of the relative momentum between the nucleons are calculated with the density operator
\begin{equation}
 \op{n}^{\mathrm{rel}}_{ST}(\vec{k}) = \sum_{i<j} \delta^{(3)}\left(\tfrac{1}{2} (\op{\vec{k}}_i-\op{\vec{k}}_j) - \vec{k} \right) \, \op{\Pi}^{ST}_{ij} \: .
\end{equation}
Results are shown in Fig.~\ref{fig:He4densQ}. The two-body momentum distributions obtained with the bare interaction can be divided into a low-momentum region up to the Fermi momentum of about 1.5 fm$^{-1}$ and a mid- to high-momentum region. The first observation is that the low-momentum region is almost independent from the SRG flow parameter. The results obtained with bare and SRG transformed density operators are very similar. We can also observe that the results for the two even channels are very similar in the low momentum region. This is in contrast to the high-momentum region. First of all the two-body densities calculated with the bare density operators are here strongly dependent on the SRG flow parameter. With increasing flow parameter the interaction gets softer and correspondingly the high-momentum components become smaller. For the largest flow parameter there are essentially no nucleon pairs with relative momenta larger than Fermi momentum. This is of course different when the SRG transformed 
density operators are used. In this case the high-momentum components observed with the bare interaction are reproduced for relative momentum larger than about 3~fm$^{-1}$. In the intermediate momentum region between 1.5 and 3~fm$^{-1}$ we find a strong $\alpha$ dependence in the $S=0,T=1$ channel and a relatively weak $\alpha$ dependence in the $S=1,T=0$ channel. Furthermore the number of pairs in the deuteron-like $S=1,T=0$ channel is larger by about a factor of ten compared to the $S=0,T=1$ channel. This can be understood by the dominance of tensor correlations in this momentum region. The strong $\alpha$-dependence in the $S=0,T=1$ channel in the mid-momentum region is again related to three-body correlations. For very high momenta above 3~fm$^{-1}$ three-body correlations are less important and the momentum distribution reflects the two-body short-range correlations coming from the repulsive core of the Argonne interaction.

The method presented in this contribution allows to study short-range correlations for all kind of interactions and also for all nuclei that are within the reach of the NCSM. It is also possible to calculate two-body densities not only as a function of the distance or the relative momentum of two nucleons but also as a function of the pair position or pair momentum.

\section*{\normalsize Acknowledgments}\vspace{-2ex}
This work was supported by the Helmholtz Alliance HA216/EMMI and by JSPS KAKENHI Grant No. 25800121.

\bibliographystyle{jpsj}
\bibliography{all}

\begin{thebibliography}{10}

\bibitem{subedi08}
R. Subedi et~al.:  Science {\bf 320} (2008) 1476.
\bibitem{forest96}
J.~L. Forest, V.~R. Pandharipande, S.~C. Pieper, R.~B. Wiringa, R. Schiavilla,
  and A. Arriaga:  Phys. Rev. C {\bf 54} (1996) 646.
\bibitem{wiringa14}
R.~B. Wiringa, R. Schiavilla, S.~C. Pieper, and J. Carlson:  Phys. Rev. C {\bf
  89} (2014) 024305.
\bibitem{alvioli13}
M. Alvioli, C. Ciofi~degli Atti, L.~P. Kaptari, C.~B. Mezzetti, and H. Morita:
  Phys. Rev. C {\bf 87} (2013) 034603.
\bibitem{ucom03}
T. Neff and H. Feldmeier:  Nucl. Phys. {\bf A713} (2003) 311.
\bibitem{ucom10}
R. Roth, T. Neff, and H. Feldmeier:  Prog. Part. Nucl. Phys. {\bf 65} (2010)
  50.
\bibitem{bogner03}
S.~K. Bogner, T.~T.~S. Kuo, and A. Schwenk:  Phys. Rept. {\bf 386} (2003) 1.
\bibitem{bogner10}
S.~K. Bogner, R.~J. Furnstahl, and A. Schwenk:  Prog. Part. Nucl. Phys. {\bf
  65} (2010) 94.
\bibitem{anderson10}
E.~R. Anderson, S.~K. Bogner, R.~J. Furnstahl, and R.~J. Perry:  Phys. Rev. C
  {\bf 82} (2010) 054001.
\bibitem{wiringa95}
R.~B. Wiringa, V.~G.~J. Stoks, and R. Schiavilla:  Phys. Rev. C {\bf 51} (1995)
  38.
\bibitem{barrett13}
B.~R. Barrett, P. Navr{\'a}til, and J.~P. Vary:  Prog. Part. Nucl. Phys. {\bf
  69} (2013) 131.
\bibitem{src11}
H. Feldmeier, W. Horiuchi, T. Neff, and Y. Suzuki:  Phys. Rev. C {\bf 84}
  (2011) 054003.
\bibitem{mitroy13}
J. Mitroy, S. Bubin, W. Horiuchi, Y. Suzuki, L. Adamowicz, W. Cencek, K.
  Szalewicz, J. Komasa, D. Blume, and K. Varga:  Rev. Mod. Phys {\bf 85} (2013)
  693.
\end{thebibliography}

\end{document}